\documentclass{epl}

\title{Four-boson scale near a Feshbach resonance}
\author{M. T. Yamashita\inst{1} \and Lauro Tomio\inst{2} \and A. Delfino\inst{3} \and T. Frederico\inst{4}}

\institute{
  \inst{1} Universidade Estadual Paulista, 18409-010, Itapeva, SP, Brazil\\
  \inst{2} Instituto de F\'\i sica Te\'orica, Universidade Estadual Paulista,
  01405-900, S\~{a}o Paulo, SP, Brazil\\
  \inst{3} Instituto de F\'\i sica, Universidade Federal Fluminense, 24210-900,
  Niter\'oi, RJ, Brazil\\
  \inst{4} Departamento de F\'\i sica, Instituto Tecnol\'ogico
de Aeron\'autica, Centro T\'ecnico Aeroespacial, 12228-900 S\~ao
Jos\'e dos Campos, SP, Brazil
}

\begin{document}

\maketitle

\begin{abstract}
We show that an independent four-body momentum scale $\mu_{(4)}$
drives the tetramer binding energy for fixed trimer energy (or
three-body scale $\mu_{(3)}$) and large scattering length ($a$).
The three- and four-body forces from the one-channel reduction of
the atomic interaction near a Feshbach resonance disentangle
$\mu_{(4)}$ and $\mu_{(3)}$. The four-body independent scale is
also manifested through a family of Tjon-lines, with slope given
by $\mu_{(4)}/\mu_{(3)}$ for $a^{-1}=0$.  There is the 
possibility of a new renormalization group limit cycle 
due to the new scale. 
\end{abstract}

Recent progress in experimental techniques, creating tunable
few-body interactions in trapped ultracold gases~\cite{cornish},
is allowing the investigation of few-body quantum phenomena, with
long wave-length triatomic properties being currently probed. In
recent experiments~\cite{WePRL03,Krcondmat06} it was confirmed the
universal properties coming from the Efimov physics~\cite{efimov}
occurring for large two-body scattering
lengths~\cite{NiPRL99,BePRL00, BrPRL01,SoPRL02,BrPRL03}. So far
experimentally, it was discovered that magnetic tunable molecular
interactions in traps allow the formation of complex molecules
like Cesium tetramers (Cs$_4$) near the scattering
threshold~\cite{chin}. One can foreseen that the physics of
four-atom systems in ultracold gases will be probed in the future
with tunable interactions.

Near a Feshbach resonance the two-atom scattering length, $a$, can
vary from very large negative values to positive values, allowing
virtual or weakly-bound dimers. The scattering length is large in
respect to the atom-atom interaction range ($r_0$), driving to the
use of concepts developed for short range interactions and halo
states~\cite{jensenrevmp}. In the limit of large $a$, the
interaction can be taken as of zero range~\cite{jackiw}. The
appearance of Thomas-Efimov states in three-boson systems is
controlled by the ratio
$r_0/|a|\rightarrow 0$~\cite{ThPR35,efimov,adh1988,mohr}.
In this exact limit, it is observed an infinite sequence of three-body
bound states~\cite{amorim97,BePRL99},
identified~\cite{albeverio,mohr} with an underlying renormalization
group limit cycle~\cite{wilson71}. The
collapse of the three-boson system when the two-body interaction
range goes to zero for fixed $a$ demands one three-body scale to
stabilize the system.

In the nuclear physics context, it was found that the $^4$He
energy is determined by the triton one, which is known as the
Tjon-line~\cite{tjon}. Once the triton energy is fitted to its
experimental value the Tjon-line gives the observed $^4$He
binding. That result could be particular to the nuclear potentials
models used in the calculations and may not be valid near a
Feshbach resonance in atomic systems.

In this letter we show that a new parameter/scale governs the
properties of four-atom systems near a Feshbach resonance. We also
discuss that the three and four-body energies can be disentangled
due to the presence of few-body forces from the one-channel
reduction. First, we discuss the atomic interaction near the
Feshbach resonance and its one-channel reduction. Second, we solve
the four-boson problem for a zero-range interaction with two and
three-boson binding energies fixed to investigate the necessity of
a new scale to determine the four-boson binding energy.

The accepted wisdom is that near a Feshbach resonance only the
two-body scale is moved through the large variations in the
atom-atom scattering length. However, surprisingly enough, there
are three and four-body potentials induced close to a Feshbach
resonance, which can disentangle the corresponding scales  and
justify a thoroughly discussion of the independence of these
scales for trapped atoms. To understand the origin of the induced
interactions, we start with the Feshbach decomposition in $P$ and
$Q$ spaces $(P+Q=1)$, with the $Q-$space representing the pair in
the potential well where it is bound, while the $P$-space
represents the pair in the lower potential well where the
continuum channel of  the ultracold regime happens. Three- and
four-body effective potentials appear when  the pair of particles
forms the bound state in the $Q-$space when interacting with
spectators. To make it concrete, we chose an example for a
three-body system $(ijk)$ where a pair $(ij)$ forms a bound state
in the well corresponding to the $Q$-space. The Hamiltonian is
$H=PHP+QHQ+ P v_{ij}Q+Q v_{ij}^\dagger P$, where the transition
potential between $P$ and $Q$ spaces is $v$. To simplify the
discussion it is assumed that the channels Hamiltonians are
$PHP=H_0+V_{ik}+V_{jk}+V_{ij}$ and $QHQ=H_0+V_{ik}+V_{jk}+
V_{ij}^Q$.  $H_0$ is the kinetic energy operator.  The gap $\Delta
E=\lim_{r\rightarrow\infty}\left(V_{ij}^Q(r) -V_{ij}(r)\right)>0$ gives the
difference in the asymptotic values of two potential wells.  The
potential $V_{ij}$ is the non-resonant part of the interaction
between $i$ and $j$, while $V_{ij}^Q$ is the second potential well
where the Feshbach resonance lies. The effective Hamiltonian
acting in the $P$-component of the wave function is
\begin{eqnarray}
H_{eff}=H_0+\sum_{r<s}V_{rs}+P v_{ij} Q G_{QQ}(E)Q v^\dagger_{ij}
P, \label{heff}
\end{eqnarray}
where the resolvent $G_{QQ}(E)=[E -H_0-  V_{ij}^Q-V_{ik}-V_{jk}
]^{-1}$. The last term in the effective Hamiltonian carries the
resonant term of the potential between  particles $(ij)$ and the
induced three-body potential, as can be understood from the
decomposition of the $Q-$channel resolvent as $G_{QQ}(E)=[E-H_0+
V_{ij}^Q]^{-1}\left[1+(V_{ik}+V_{jk})G_{QQ}(E) \right]$. The
resonant interaction between particles ${ij}$ is given by $P
v_{ij} Q [E-H_0-V_{ij}^Q ]^{-1}Q v^\dagger_{ij} P$, when it is
accounted the dominance of the projection over the bound state of
the potential $V_{ij}^Q $ near the scattering threshold. The
three-body connected part of the remaining term corresponds to a
three-body interaction, with intensity clearly depending on the
position of the Feshbach resonance. Such, reasonings are easily
extended to four particles. Therefore, in principle not only the
scattering length can be tuned near a Feshbach resonance but also
the three and four-body binding energies. Turning off the
excitation of the Feshbach resonance in the presence of spectator
particles, the formalism is reduced to the one discussed in
Ref.~\cite{jonsell}.

In short, near a Feshbach resonance the induced one-channel 
three and four boson forces can drive independently the corresponding
physical scales. Therefore four-body observables, like four-boson
recombination rates or atom-trimer or dimer-dimer scattering
lengths can exhibit correlations not constrained by  one
low-energy s-wave three-boson observable and $a$, if a four-boson
scale really exists.

We begin our discussion of the relevant scales in the few-boson
system reminding the form of the regulated trimer bound-state
integral equation, in units of $\hbar=m=1$ ($m$ is the boson
mass), which is written as
\begin{eqnarray}
f(q)=
 \frac{\pi^{-1}}{-a^{-1}+\sqrt{|E_3|+\frac34q^2} }
\int_0^\infty k^2dk~f(k) \int^1_{-1}dz \left[\frac{1}{|E_3|+q^2+k^2+qkz}
-\frac{1}{\mu_{(3)}^2+q^2+k^2+qkz}\right] , \label{trimer}
\end{eqnarray}
where the second term in Eq.~(\ref{trimer}) brings the physical
scale, $\mu_{(3)}\sim r_0^{-1}$, to the three-boson
system~\cite{AdPRL95} producing a finite ground-state energy and
avoiding the Thomas collapse~\cite{ThPR35}.

The sensitivity of  three-boson s-wave observables to the
short-range part of the interaction in weakly bound systems is
parameterized through the value of the trimer binding energy which
corresponds to the  regularization scale $\mu_{(3)}$. The s-wave
three-boson observables are strongly correlated to the trimer
energy and scales in the general universal
form~\cite{amorim97,virtual}:
\begin{eqnarray}
{\cal O}_3(E, E_3, a)=|E_3|^\eta {\cal
F}_3\left(\frac{E}{E_3},a\sqrt{|E_3|}\right) \ , \label{obs3}
\end{eqnarray}
where ${\cal O}_3$ can represent an observable at an energy $E$ or
an excited trimer energy (the dependence on $E$ does not appear in
this last case). The exponent $\eta$ gives the correct dimension
to ${\cal O}_3 $. Eq.~(\ref{trimer}) is renormalization group (RG)
invariant with its kernel being a solution of a non-relativistic
Callan-Symanzik differential equation~\cite{nrcs} as function of a
sliding $\mu_{(3)}$. In that way $E_3$ and three-body observables
are independent of the subtraction point (see Ref.~\cite{virtual}
for a discussion of the RG invariance in three-body systems).

In a four-boson system with short-range interactions one may
be tempted to think in an independent momentum scale, which is still
under discussion~\cite{fedorovjensen2002,platter}. In the
hyperspherical expansion method used in Ref.~\cite{fedorovjensen2002,jensenrevmp}
it is enough to fix the scattering length, effective range and
shape parameter in the limit of zero-range interaction to stabilizes
the N-boson system. On the other hand within effective field
framework it is enough a repulsive three-body force that stabilizes
the trimer to allow finite values of the tetramer binding energy~\cite{platter}.

In our strategy to verify the necessity of a new four-boson scale,
the dimer binding energy is set to zero while the trimer
ground-state energy is kept fixed.  Within these assumptions, to
allow finite results for the tetramer energy, the
Faddeev-Yakubovsky (FY) equations~\cite{CiPRC98} should be
regularized with the insertion of  momentum cutoffs or subtraction
scales. From our results for the binding energies in the limit of
a zero-range interaction, it emerges the necessity of a new scale
in the four-boson system, which depends crucially on the four-body
scale, the subtraction point $\mu^2_{(4)}$. The existence of an 
independent four body scale can be in principle checked by measuring 
$s-$wave observables for three and four atoms near a Feshbach resonance. 
If the three and four-body scales can be made truly independent 
due to multi-atom forces, one can distinguish from the case that
such scales are equal considering the correlation 
between two observables.

In the next, we show the systematical inclusion of the three and
four-body subtraction points in the unregulated FY equations.
First, it is natural to require $\mu_{(3)}$ as the subtraction
scale of the embedded three-boson system in the four-body one,
otherwise the threshold for the four-boson energy would be
different from the ground-state trimer energy. Second, the
four-boson scale  or the new subtraction point is introduced in
the set of FY equations, and the tetramer binding energy is
calculated for different values of $\mu_{(4)}$.
\begin{figure}
\onefigure{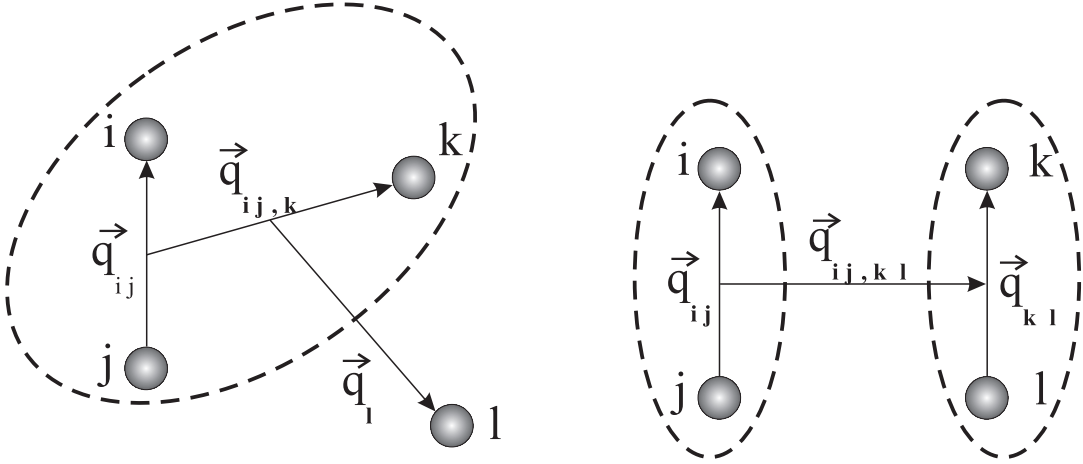}
\caption{Jacobi coordinates for the Faddeev-Yakubovsky
amplitudes. $K$-type, left-side, and $H$-type, right side.}
\label{kh}
\end{figure}

The dynamics of a four-boson system can be described in terms of
the Faddeev-Yakubovsky amplitudes~\cite{CiPRC98}, $K$ and $H$,
represented schematically in Fig.~(\ref{kh}). The amplitude
$K_{ij,k}^l$ corresponds to the partition in a three-body
subsystem formed by particles, $ijk$, and the spectator one, $l$,
while $H_{ij,kl}$ describes the partition in two subsystems formed
by the pairs $ij$ and $kl$. There are 18 independent FY
amplitudes, 12 $K$-type and 6 $H$-type. The bound state FY
amplitudes satisfy:
\begin{eqnarray}
|K_{ij,k}^l\rangle&=&G_0t_{ij}(E-E_{ij,k}-E_l)\big[|K_{ik,j}^l\rangle+
|K_{jk,i}^l\rangle \nonumber \\
&+&|K_{ik,l}^j\rangle+|K_{jk,l}^i\rangle+|H_{ik,jl}\rangle
+|H_{jk,il}\rangle\big], \label{kf} \\
|H_{ij,kl}\rangle&=&G_0t_{ij}(E-E_{ij,kl}-E_{kl})\big[|K_{kl,i}^j\rangle
+|K_{kl,j}^i\rangle+ |H_{kl,ij}\rangle\big],
\label{hf}
\end{eqnarray}
where the four-body free resolvent is $G_0=(E- H_0)^{-1}$ ($ H_0$ is
the kinetic energy operator) and $t_{ij}$ is the two-body T-matrix.
The kinetic energies for the Jacobi momenta for identical particles
of unit mass are $E_{ij,k}=\frac34{ q}_{ij,k}^2$, $E_l=\frac23{
q}_{l}^2$, $E_{ij,kl}=\frac12{ q}_{ij,kl}^2$ $E_{kl}={ q}_{kl}^2$.
The total wave function of the four-body system  is given by the sum
$|\Psi\rangle=\sum_{i<j}|F_{ij}\rangle$, where the Faddeev
components $|F_{ij}\rangle=G_0V_{ij}|\Psi\rangle$ can be written in
terms of the FY amplitudes as
$|F_{ij}\rangle=|K_{ij,k}^l\rangle+|K_{ij,l}^k\rangle+|H_{ij,kl}\rangle.$

The renormalized two-body T-matrix for a zero-range interaction is
given by $t_{ij}(x)= |\chi_{ij}\rangle\tau(x)\langle\chi_{ij}| $
with $\tau^{-1}(x)=2\pi^{2}\left[a^{-1}-\sqrt{-x}\right]$  and
$\langle \vec{q}_{ij}|\chi_{ij}\rangle=1$. The structure of the FY
equations is simplified for the zero-range interaction once the
separation $|K_{ij,k}^l\rangle=G_0|\chi_{ij}\rangle |{\cal K
}_{ij,k}^l\rangle$ and $|H_{ij,kl}\rangle=G_0|\chi_{ij}\rangle
|{\cal H}_{ij,kl}\rangle$ is recognized in Eqs.(\ref{kf}) and
(\ref{hf}). The FY reduced amplitudes, ${\cal K }$ and ${\cal H
}$, satisfy a coupled set of integral equations which needs
regularization. One recognizes that the resolvent of the immersed
three-boson subsystem should carry the regularization scale
$\mu_{(3)}$ to avoid the Thomas collapse of the system. This is
done in analogy with Eq.~(\ref{trimer}), where the momentum
integration is regularized by a subtracted Green's function in a
subtraction point $\mu_{(3)}^2$. In the l.h.s. of
Eq.~(\ref{subs1}) one sees the correspondence with the three-boson
equation written as $\tau^{-1}(e^{(1,3)}_{2} )|{\cal K
}_{ij,k}^l\rangle-  {\cal G}^{(3)}_{ij;ik} |{\cal K}
_{ik,j}^l\rangle -  {\cal G}^{(3)}_{ij;jk}|{\cal K}
_{jk,i}^l\rangle =0$. However, other terms are present in the
four-body equations which require regularization.

We introduce the scale $\mu_{(4)}$ such that the four-body free
Green's function $G_0(E_4)$ is  substituted by
$G_0(E_4)-G_0(-\mu_{(4)}^2)$ in a direct generalization of
Eq.~(\ref{trimer}), as suggested by ~\cite{AdPRL95}:
\begin{eqnarray}
\nonumber \tau^{-1}(e^{(1,3)}_{2} )|{\cal K }_{ij,k}^l\rangle-
 {\cal G}^{(3)}_{ij;ik} |{\cal K} _{ik,j}^l\rangle
-  {\cal G}^{(3)}_{ij;jk}|{\cal K} _{jk,i}^l\rangle = \\
  {\cal G}^{(4)}_{ij;ik}\big[ |{\cal K}_{ik,l}^j\rangle +
|{\cal H}_{ik,jl}\rangle\big]+  {\cal G}^{(4)}_{ij;jk} \big[|{\cal
K}_{jk,l}^i\rangle + |{\cal H}_{jk,il}\rangle\big], \label{subs1}
\\
 \tau^{-1}(e^{(2,2)}_{2}) |{\cal H} _{ij,kl}\rangle=  {\cal
G}^{(4)}_{ij;kl}\big[ |{\cal K }_{kl,i}^j\rangle + |{\cal
K}_{kl,j}^i\rangle +|{\cal H}_{kl,ij}\rangle\big] , \label{subs2}
\end{eqnarray}
where $e^{(1,3)}_{2}=E-E_{ij,k}-E_l$ and
$e^{(2,2)}_{2}=E-E_{ij,kl}-E_{kl}$ are
 the two-body subsystem energies in the (1,3) and (2,2) partitions,
 respectively. The projected Green's function operators are
$ {\cal
G}^{(N)}_{ij;ik}:=\langle\chi_{ij}|{G}^{(N)}_{0}|\chi_{ik}\rangle $
with $N$ equal 3 or 4, with the subtracted Green's functions given
by $G^{(3)}_0=[E-H_0]^{-1}-[-\mu_{(3)}^2- H_0]^{-1}$ and
$G^{(4)}_0=[E- H_0]^{-1}-[-\mu_{(4)}^2- H_0]^{-1}$. Maintaining the
two- and the three-body scales fixed, the four-body scale can be
moved and its consequences for the tetramer binding investigated.

The four-boson integral equations for the reduced FY amplitudes are
derived from the projection in Eqs.~(\ref{subs1}) and
(\ref{subs2}) in Jacobi relative momenta.
The ground state of the four-boson system has total angular momentum
zero, therefore the spectator functions depends only on the modulus
of the momenta. The integral equations are analogous to the ones
shown in Ref.~\cite{platter} (without three-body forces),
but with the four-body Green's function substituted by their
subtracted form as given in Eqs.~(\ref{subs1}) and (\ref{subs2}).
The numerical procedure followed Ref.\cite{x}.

The momentum scales in Eqs.~(\ref{subs1}) and (\ref{subs2}) are
$a^{-1}$, $\mu_{(3)}$ and $\mu_{(4)}$, consequently the tetramer
binding energy for $a^{-1}=0$ depends on the momentum scales as
\begin{eqnarray}
E_4/\mu^2_{(3)}~= {\cal {E} }_4\left(\mu_{(4)}/\mu_{(3)}\right).
\label{sca4b}
\end{eqnarray}

In  Table I, we present results for the tetramer ground state
binding energy for $|a|=\infty$ and considering a fixed
$\mu_{(3)}$, such that we take $\mu_{(3)}$ as our momentum
unit.  From the numerical solution of Eqs.~(\ref{subs1}) and (\ref{subs2})
we obtain ${\cal {E} }_4/\mu_{(3)}^2$ for different values of
$\mu_{(4)}/\mu_{(3)}$.  The trimer binding energy is fixed and
given by the solution of Eq.~(\ref{trimer}),
$E_3 =-0.00927 \mu_{(3)}^2$~\cite{virtual}.
We clearly observe in this table the dependence of $E_4$ on
$\mu_{(4)}$ (for fixed $\mu_{(3)}$). As $\mu_{(4)}/\mu_{(3)}$ is
varied from 1 to 20, the ratio $E_4/E_3$ changes from 5 to about 78
(considering two digits). In order to have
${\cal {E} }_4$ independent on the regularization scale
$\mu_{(4)}$, one should require that for $\mu_{(4)}/\mu_{(3)}>>1$
the four-body scale would vanish as a physical scale.
This is not observing in our results.

In view of the scaling relation Eq.~(\ref{sca4b}), the results
presented in the table are valid for any $\mu_{(3)}$.
Using that  $E_3 =-0.00927 \mu_{(3)}^2$ in Eq.~(\ref{sca4b})
we obtain a  family of Tjon lines, where each one corresponds to
a given value of $\mu_{(4)}/\mu_{(3)}$. From this result arises quite
naturally the main claim of our work: the existence of a four-body
scale, which is manifested through a family of Tjon-lines.
The line slope is given by the ratio $\mu_{(4)}/\mu_{(3)}$.

\begin{table}
\caption{Four-boson binding energy for $E_2=0$ and fixed
$\mu_{(3)}$. The three-boson energy is $E_3=-0.00927\mu^2_{(3)}$~\cite{virtual}. }
\label{e4}
\begin{center}
\begin{tabular}{|c|ccccccc|} \hline
$\mu_{(4)}/\mu_{(3)}$ & 1 & 2 & 4 & 7 &10 & 15& 20 \\ \hline
$E_4/E_3$ & 5.0 & 7.8 & 13 &22 &29 & 51 & 78 \\ \hline
$E_4/\mu^2_{(3)}$ &-0.046  &  -0.072  & -0.124 &-0.20 & -0.27  &-0.47&  -0.72 \\
\hline
\end{tabular}
\end{center}
\end{table}

For equal three and four-body scales, i.e.,
$\mu_{(4)}=\mu_{(3)}=1$, the tetramer binding energy is
$E_4\simeq5~E_3$. The Tjon line~\cite{tjon} is
$E_\alpha=4.72(E_t+2.48)$~MeV ($E_\alpha$ is the $^4$He binding
energy and $E_t$ the triton one), with the slope
approached by our calculation. Also, it was recently found in
Ref.~\cite{platter} that $E_4$ scales as $\sim 5~E_3$, for equal
bosons interacting through a zero-range two-body force, with a
repulsive three-body potential needed to stabilize the trimer
ground state energy against collapse. Our result for
$\mu_{(4)}=\mu_{(3)}$ agrees with both Refs.~\cite{tjon} and
\cite{platter}. The physical reason for this agreement is
explained in the next two paragraphs.

In the nuclear case, the nucleon-nucleon interaction is strongly
repulsive at short distances and therefore the probability to have
four nucleons simultaneously inside a volume $\sim r_0^3$ is quite
small. For other potentials, like separable ones, parameterizing
the nuclear force range, a similar result is found~\cite{gibson},
even without a short range repulsion due to the non-locality of
the interaction. In these cases, presumably the four-nucleon scale
itself has much less opportunity to be evidenced, i.e., the three-
and four-body scales correspond to the same short-range physics,
which in our calculation come from $\mu_{(4)}=\mu_{(3)}$.

In Ref.~\cite{platter}, a three-body force is used to stabilize
the shallowest three-body state, against the variation of the
cut-off. Their results show that the Tjon line is reproduced when
the three-body force is repulsive. Within this picture, the
contribution of the three-body interaction in a configuration
where the four bosons interact simultaneously is about four times
more repulsive than having three-bosons in the force range and the
other outside this range. Therefore, the dynamics of the
short-range part of the interaction is manifested in the
four-boson system mainly through small three-boson configurations,
this should also qualitatively corresponds to the particular case
$\mu_{(4)}=\mu_{(3)}$.

Also, we should note that, in Ref.~\cite{petrov}, considering
the three-boson problem near a {\it narrow} Feshbach resonance,
Petrov concludes that three-body observables depend only on the
resonance width (which implies in an effective range of the order
of $\mu_{(3)}^{-1}$) and the scattering length. One may suspect that
his conclusion can also have consequences to four and more
particle observables. However, we would like to stress that,
within our present coupled-channel formalism, three and more body
forces may appear, giving more freedom to the three and four atom
systems. If one sticks to the Petrov's assumption, than the three
and four body scales are equal ($\mu_{(3)}=\mu_{(4)}$) and no
independent behavior can be verified in practice.

The general scaling form of  s-wave three-boson observables,
Eq.~(\ref{obs3}), can be generalized to four-bosons with the
dynamics parameterized by the scattering length, trimer and
tetramer binding energies. A s-wave four-boson observable will be
strongly correlated to $a$, $E_3$ and $E_4$:
\begin{eqnarray}
{\cal O}_4(E, E_4,E_3, a)=|E_4|^\eta {\cal
F}_4\left(\frac{E}{E_4},\frac{E_3}{E_4},a\sqrt{|E_4|}\right) \ ,
\label{obs4}
\end{eqnarray}
where ${\cal O}_4 $ represents either an observable at energy $E$
or an excited tetramer energy (the dependence on $E$ does not
appear in the last case). The exponent $\eta$ gives the correct
dimension to ${\cal O}_4 $. There is the 
possibility of a new renormalization group limit cycle 
due to the new scale, and for $\mu_{(4)}\rightarrow \infty$ one may 
speculate that the correlation
given by Eq.~(\ref{obs4}) would approach a universal function.

In summary, by studying a four-boson system with a zero-range
interaction, we conclude that a four-body scale is necessary when
the two and three-boson scales are fixed.  Extrapolating this
result to more particles, we suggest that one new experimental
information is required for each new particle added to the system
near a Feshbach resonance, to include effects from many-body
forces from the one-channel reduction. In the particular case,
when only two-body forces are present, all scales are equal and
can be for example associated with the effective range. However,
as more accuracy is needed to describe the physics near a Feshbach
resonance a distinction between 3 and 4 body scales must be
introduced.

\acknowledgments
We thank the Brazilian agencies FAPESP and CNPq for partial
support.

\end{document}